# FUNDAMENTAL AND VORTEX DISSIPATIVE QUADRATIC SOLITONS SUPPORTED BY SPATIALLY LOCALIZED GAIN

Valery E. Lobanov[1], Aleksey A. Kalinovich[2], Olga V. Borovkova[2], and Boris A. Malomed[3,4]

[1]Russian Quantum Center, Skolkovo 143026, Russia

[2]Faculty of Physics, Lomonosov Moscow State University, Moscow 119991, Russia

[3]Department of Physical Electronics, School of Electrical Engineering, Faculty of Engineering, Tel Aviv University, Tel Aviv 69978, Israel

[4]Instituto de Alta Investigación, Universidad de Tarapacá, Casilla 7D, Arica, Chile

## Abstract

We consider settings providing the existence of stable two-dimensional (2D) dissipative solitons with zero and nonzero vorticity in optical media with quadratic ($\chi^{(2)}$) nonlinearity. To compensate the spatially uniform loss in both the fundamental-frequency (FF) and second-harmonic (SH) components of the system, a strongly localized amplifying region ("hot spot", HS), carrying the linear gain, is included, acting on either the FF component or the SH one. In both cases, the Gaussian radial gain profile supports stable fundamental dissipative solitons pinned to the HS. The structure of existence and stability domains for the 2D solitons is rather complex. They demonstrate noteworthy features, such as bistability and spontaneous symmetry breaking. A ring-shaped gain profile acting on the FF component supports stable vortex solitons, with the winding number up to 5, and multipoles. Nontrivial transformation of vortex-soliton profiles upon either growth of the gain value or stretching of the gain profile is observed. © 2021 The Authors

## 1. Introduction

Generation of different types of solitary waves or solitons is one of the central topics in nonlinear optics [1]. Dissipative effects, which are inevitable in many practically important settings, have initiated a new direction of the research for optical solitons [2-10]. As a result, spatial solitary waves, self-trapped in the transverse directions, have been found as robust states not only in conservative but also in dissipative optical media. In the latter case, the conditions of the balance between the diffraction and self-focusing, as in lossless media, and also between the loss and the gain, must hold simultaneously [2-4]. Dissipative spatial solitons have been studied in various settings [4], including semiconductor amplifiers [11], lasers with saturable gain and absorption [2,12], Bose-Einstein condensates [13,14], models described by the cubic–quintic complex Ginzburg–Landau (CGL) equation [15–20], etc.

Relevant possibilities for the generation of localized nonlinear dissipative patterns have been demonstrated in setups combining spatially localized gain ("hot spots", HSs) and uniform linear or nonlinear losses (see Refs. [21, 22] for review). Detailed analytical and numerical analysis has been performed for one- [23-37], two- [38-43], and even three-dimensional (for spatiotemporal localization of the gain profile) [44, 45] realizations of this model. Besides fundamental solitons, ring-shaped gain profiles have been found to support stable vortex solitons [40-43].

Solitonic structures have been studied as well for the systems with periodic distributions of the gain and loss [46], and in discrete systems (lossy lattices) with the gain applied at a particular site of the lattice [47-49]. Stable dissipative solitons have been also found in the 1D system with the spatially uniform linear gain and nonlinear loss whose strength grows from the center to periphery of the system at any rate faster than the distance from the center [50]. A combination of symmetrically placed and mutually balanced localized gain and loss elements supports localized states featuring the *PT*- (parity-time) symmetry [51-54].

The above-mentioned works addressed media with uniform or localized Kerr (cubic) nonlinearity, while dissipative solitons in materials with quadratic ($\chi^{(2)}$) nonlinearity have been significantly less studied. An advantage of such materials is very strong nonlinearity, in comparison to their Kerr counterparts, hence the respective size of the experimental samples and laser-beam powers required for the observation of solitons can be reduced significantly [55-60]. However, a majority of actual materials with high values of the $\chi^{(2)}$ coefficient (e.g., organic monocrystals and semiconductors, such as InAs, InSb, GaSb) also feature high absorption level at optical wavelengths. Therefore, the concept of spatially localized gain, which provides compensation of intrinsic absorption, may be very effective in facilitating soliton generation in quadratic media. Previously, dissipative $\chi^{(2)}$ solitons have been studied in models of spatially uniform optical cavities [61-63] and ones with *PT*-symmetric [64-66] and more general [67] localized potentials. Thus far, the use of localized gain in quadratic media was considered solely in a 1D system [68].

Here we study the existence and stability of 2D fundamental (zero-vorticity) and vortex solitons supported by spatially localized linear gain acting on the single component (fundamental-frequency (FF) or second-harmonic (SH) one) in $\chi^{(2)}$ systems with uniform linear losses. In both cases, of the FF or SH amplification, a Gaussian gain profile supports stable fundamental $\chi^{(2)}$ solitons pinned to the gain region. Shapes of the existence and stability domains for the solitons in the parameter space, produced by numerical analysis, can be rather complex, and stable solitons demonstrate noteworthy features, such as bistability and spontaneous symmetry breaking. Further, a ring-shaped gain profile acting on the FF component supports stable vortex solitons with winding numbers (topological charges) up to 5, and multipoles. Nontrivial transformations of the vortex solitons following either growth of the gain or expansion of the HS profile are observed.

The rest of the paper is organized as follows. The model is formulated in Section 2. Systematic results demonstrating the existence and stability of fundamental (zero-vorticity) solitons supported by the gain applied to the FF and SH components are reported, severally, in Sections 3 and 4. In Section 5, the analysis is extended for vortex states supported by the FF gain (stable vortex solitons were not found in the system with the gain acting in the SH component). Stable vortices are found with winding numbers up to $m = 5$. In cases when they are unstable against azimuthal perturbations, stable modes in the form of rotating *azimuthons* emerge spontaneously. Also stable may be higher-order circular multipoles (e.g., ones composed of 12 segments with alternating signs). The paper is completed by Section 6.

## 2. The model

First, we introduce the system with the two-dimensional HS acting on the FF component. It is based on the system of coupled scaled equations for local amplitudes of the FF ($q_1$) and SH ($q_2$) waves in the spatial domain, under conditions for the type-I (degenerate, i.e., two-wave) $\chi^{(2)}$ interaction [68] in the presence of the dissipation and localized gain:

$$\begin{cases} i\dfrac{\partial q_1}{\partial \xi} = -\dfrac{1}{2}\left(\dfrac{\partial^2 q_1}{\partial \eta^2} + \dfrac{\partial^2 q_1}{\partial \zeta^2}\right) - q_1^* q_2 \exp\left(-i\beta\xi\right) + i\gamma_1\left(\eta, \zeta\right) q_1, \\ i\dfrac{\partial q_2}{\partial \xi} = -\dfrac{1}{4}\left(\dfrac{\partial^2 q_2}{\partial \eta^2} + \dfrac{\partial^2 q_2}{\partial \zeta^2}\right) - q_1^2 \exp\left(i\beta\xi\right) - i\gamma_{20} q_2. \end{cases} \quad (1)$$

where $\eta$, $\zeta$ and $\xi$ are the normalized transverse coordinates and propagation distance, respectively. The radially-symmetric gain-loss profile, with strength $a_1 > 0$, spatial width $W$ of the HS, acting in the FF component, and the uniform FF loss $\gamma_{10} > 0$, is

$$\gamma_1\left(\eta,\zeta\right)=a_1\exp\left(-\left(r-R\right)^2/W^2\right)-\gamma_{10}. \qquad (2)$$

Here, $r=\sqrt{\eta^2+\zeta^2}$ is the radial coordinate, and radial profiles (2) with $R=0$ and $R>0$ represent, respectively, the Gaussian and ring-shaped HS profiles. Further, $\gamma_{20}>0$ is the homogeneous loss acting on the SH, and $\beta$ is the wavenumber mismatch. The localized gain profiles represented by Eq. (2) may be realized by suitable pump beams, or by shaping the concentration of active centers distributed in the optical medium.

It is relevant to mention that, in the case of relatively large values of mismatch $|\beta|$ and SH loss coefficient $\gamma_{20}$, one can apply the cascading approximation [57] to the second equation in system (1), which yields $q_2\approx\left(\beta-i\gamma_{20}\right)^{-1}e^{i\beta\xi}q_1^2$. Inserting this in the first equation leads to an effective cubic CGL equation with nonlinear loss:

$$i\frac{\partial q_1}{\partial \xi}=-\frac{1}{2}\left(\frac{\partial^2 q_1}{\partial \eta^2}+\frac{\partial^2 q_1}{\partial \zeta^2}\right)-\frac{\beta+i\gamma_{20}}{\beta^2+\gamma_{20}^2}\left|q_1\right|^2 q_1+i\gamma_1\left(\eta,\zeta\right)q_1. \qquad (3)$$

This equation gives rise to families of dissipative solitons pinned to the HS [21, 22], with the nonlinear-loss term providing additional stabilization.

The system with the localized gain acting at the SH is adopted in the following form:

$$\begin{cases} i\dfrac{\partial q_1}{\partial \xi}=-\dfrac{1}{2}\left(\dfrac{\partial^2 q_1}{\partial \eta^2}+\dfrac{\partial^2 q_1}{\partial \zeta^2}\right)-q_1^* q_2\exp\left(-i\beta\xi\right)-i\gamma_{10}q_1, \\[4mm] i\dfrac{\partial q_2}{\partial \xi}=-\dfrac{1}{4}\left(\dfrac{\partial^2 q_2}{\partial \eta^2}+\dfrac{\partial^2 q_2}{\partial \zeta^2}\right)-q_1^2\exp\left(i\beta\xi\right)+i\gamma_2\left(\eta,\zeta\right)q_2. \end{cases} \qquad (4)$$

where $\gamma_{10}$ is the spatially uniform FF loss coefficient, while the spatial profile of the gain at the SH is taken as in Eq. (2), *viz.*,

$$\gamma_2\left(\eta,\zeta\right)=a_2\exp\left(-\left(r-R\right)^2/W^2\right)-\gamma_{20}. \qquad (5)$$

# 3. Fundamental solitons supported by the localized gain applied at the fundamental-frequency harmonic

First, we considered the Gaussian gain profile with $R = 0$ and fixed width and strength parameters, $W = 1.0$ and $\gamma_{10} = 1.0$. Stationary radially symmetric solutions with real propagation constant $b$ are looked for as

$$q_1 = w_1(r)\exp(ib\xi), \quad q_2 = w_2(r)\exp(i(2b+\beta)\xi), \tag{6}$$

where $w_{1,2}(r)$ are complex functions satisfying the following radial equations:

$$\begin{cases} \dfrac{1}{2}\left(\dfrac{\partial^2 w_1}{\partial r^2} + \dfrac{1}{r}\dfrac{\partial^2 w_1}{\partial r}\right) - bw_1 + w_1^* w_2 - i\gamma_1(r)w_1 = 0, \\[3mm] \dfrac{1}{4}\left(\dfrac{\partial^2 w_2}{\partial r^2} + \dfrac{1}{r}\dfrac{\partial^2 w_2}{\partial r^2}\right) - (2b+\beta)w_2 + w_1^2 + i\gamma_{20}w_2 = 0, \end{cases} \tag{7}$$

with * standing for the complex conjugate. The states are characterized by their FF and SH powers, $U_{1,2} = \iint |q_{1,2}(x,y)|^2 dxdy$, the total power (alias energy flow, or the Manley-Rowe invariant [69]), $U = U_1 + U_2$, being the dynamical invariant of the system in the absence of the loss and gain. Another dynamical invariant is the $\xi$ component of the angular momentum,

$$M = M_1 + M_2, M_n = \dfrac{i}{n}\iint q_{1,2}^*\left(\zeta\dfrac{\partial q_{1,2}}{\partial \eta} - \eta\dfrac{\partial q_{1,2}}{\partial \zeta}\right)d\eta d\zeta, \, n = 1,2. \tag{8}$$

In the presence of the gain and loss, stationary solutions must satisfy the power-balance condition:

$$\dfrac{dU}{d\xi} = 2\left[2\pi a_1\int_{-\infty}^{+\infty}\exp\left(-\dfrac{r^2}{W^2}\right)|w_1(r)|^2 rdr - \gamma_{10}U_1 - \gamma_{20}U_2\right] = 0. \tag{9}$$

First, radially symmetric numerical solutions of Eq. (7) were constructed by means of the relaxation method and the soliton existence domain was identified. In particular, numerically constructed dissipative $\chi^{(2)}$ solitons (see a typical example in Fig. 1) exist if the strength of the HS gain exceeds a certain minimum value, $a_1 > a_{10}$ (see Fig. 2), where $a_{10}$ is the HS amplitude supporting linear (small-amplitude) gain-guided modes [70,71], and the mismatch parameter is not too large, $\beta < \beta_{cr}(a_1, \gamma_{20})$, see Fig. 3. For the present values of parameters, $a_{10} \approx 2.65$.

Naturally, the value of $a_{10}$ decreases with the growth of the HS width $W$, as broader HS provides larger overall gain.

The existence and stability of the stationary solutions, produced by the numerical solution of Eq. (7), were also examined via the direct propagation governed by Eq. (1) with the input in the form of the FF Gaussian beam, using the standard split-step fast-Fourier-transform algorithm. The stability of the soliton solutions was then tested by imposing random perturbations (additive and/or multiplicative) to the solutions and simulating their subsequent evolution up to $\xi \sim 10^3$. Solitons that kept their shape in the course of the propagation were classified as stable ones. Figure 1 shows a typical profile of a stable dissipative soliton. Note that stable soliton solutions located at the maximum of the gain profile exhibit a nontrivial radial phase distribution providing energy flow from the HS to the loss-dominated periphery.

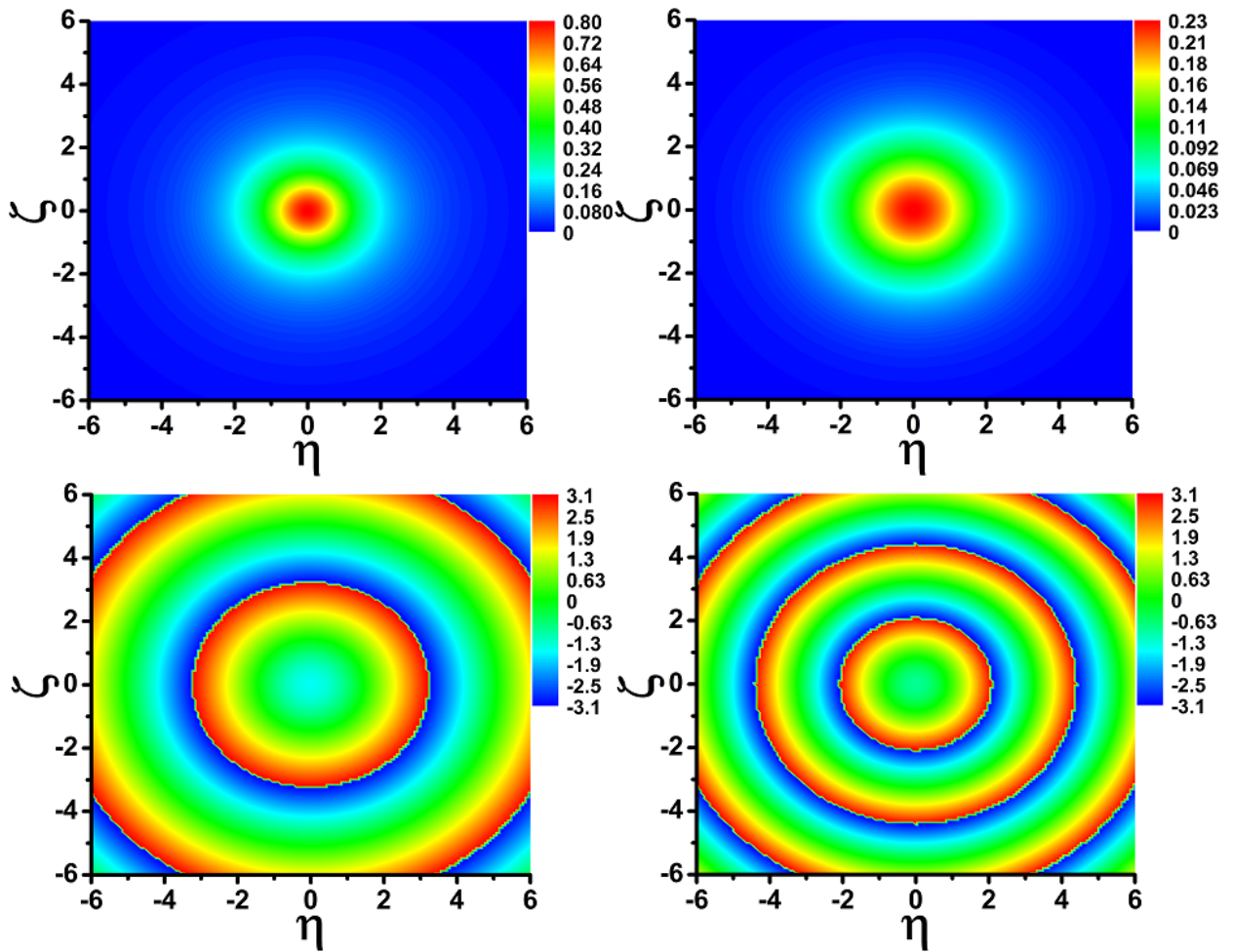

**Fig. 1.** The FF and SH components (left and right columns, respectively) in a typical 2D fundamental dissipative quadratic soliton driven by the gain at the FF component: The absolute value of the fields (upper panels) and their phase (bottom panels), produced by the numerical solution of Eq. (6) at $a_1 = 3.0$, $\beta = -1.0$, $W = 1.0$, and $\gamma_{10} = \gamma_{20} = 1.0$. Weak "jitter" in phase distributions is a manifestation of a finite step of the transverse grid used for the calculations. All quantities are plotted in dimensionless units.

Interestingly, the stable solitons can exist not only at negative mismatches corresponding to the defocusing cascading nonlinearity [51] but also in some range of positive values of $\beta$, see Fig. 4. However, while at $\beta < 0$ the solitons are found at any gain strength exceeding the critical value, at $\beta > 0$ there is a certain range of values of the gain coefficient where solitons do not exist even at $a_1 > a_{10}$ [the existence discontinuity is clearly seen in the right panel of Fig. 2]. Inside of the existence domain, the power of the soliton components grows monotonically with the increase of the gain coefficient.

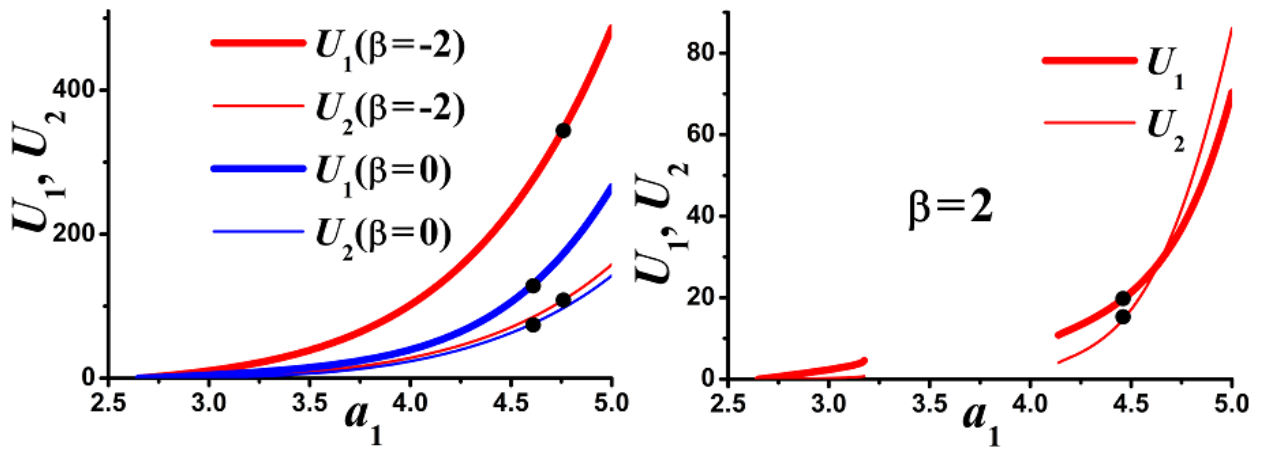

**Fig. 2.** The powers of the soliton components vs. the FF-gain coefficient, $a_1$, at different values of mismatch $\beta$. Black dots indicate boundaries of the stability domains (solitons are stable to the right of these points). In all cases $W = 1.0$, $\gamma_{10} = \gamma_{20} = 1.0$. All quantities are plotted in dimensionless units.

Existing at large absolute values of the negative mismatch, the solitons do not exist at large positive values of $\beta$, see the right panel in Fig. 3. At the threshold value of $\beta$, which depends on the gain strength, the tangential line to the $U_{1,2}(\beta)$ curve becomes vertical, while $U_{1,2}$ remain finite (see Fig. 3). At $\beta$ exceeding the threshold value, soliton-like inputs exhibit blowup (not shown here in detail). The powers of the soliton components, especially of the FF one, tend to grow with the decrease of $\beta$ (see the right panel in Fig. 3), and the soliton profile widens.

The stability analysis demonstrates that the dissipative solitons become unstable when the gain strength exceeds a certain critical value, $a_1 > a_{1cr}(\beta, \gamma_{20})$, which slightly depends on the mismatch. Unstable solitons spontaneously transform into some breather-like modes (see the left panel in Fig. 4). As it is shown in the left panel of Fig. 4, the power of stable solitons remains constant in the course of the propagation (as shown by the blue line), while the power of unstable ones oscillates in the course of the propagation, covering the red stripe. The width of the stability domain increases with the growth of the SH loss strength, $\gamma_{20}$ (see Fig. 4, right panel). Thus, stable solitons exist in a finite range of the gain value, and below the critical mismatch.

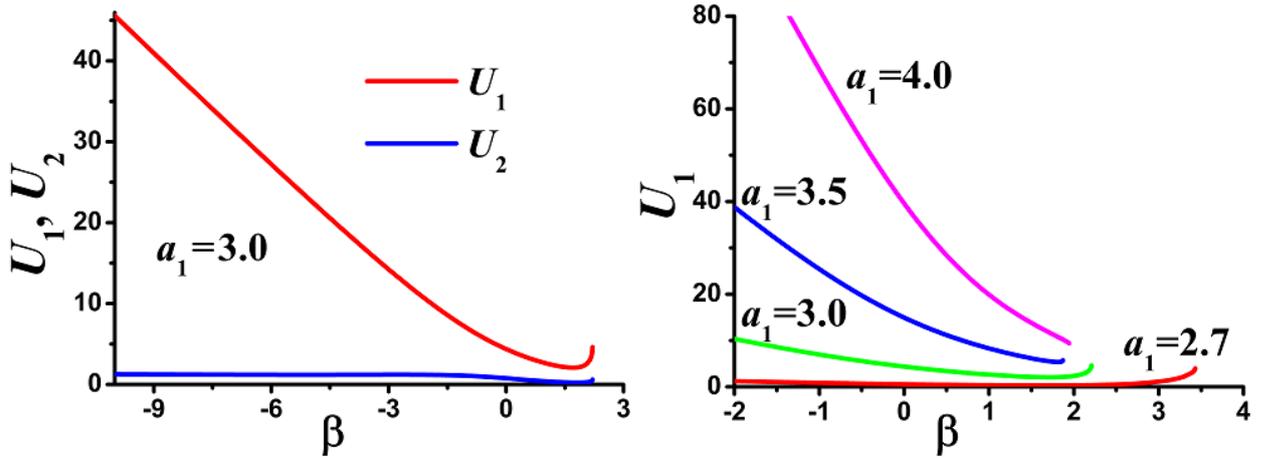

**Fig. 3.** Left: The power of the components of the FF-driven solitons vs. $\beta$ for $a_1 = 3.0$. Right: The power of the FF component vs. $\beta$ at different values of $a_1$. In all cases $W = 1.0$, $\gamma_{10} = \gamma_{20} = 1.0$. All quantities are plotted in dimensionless units.

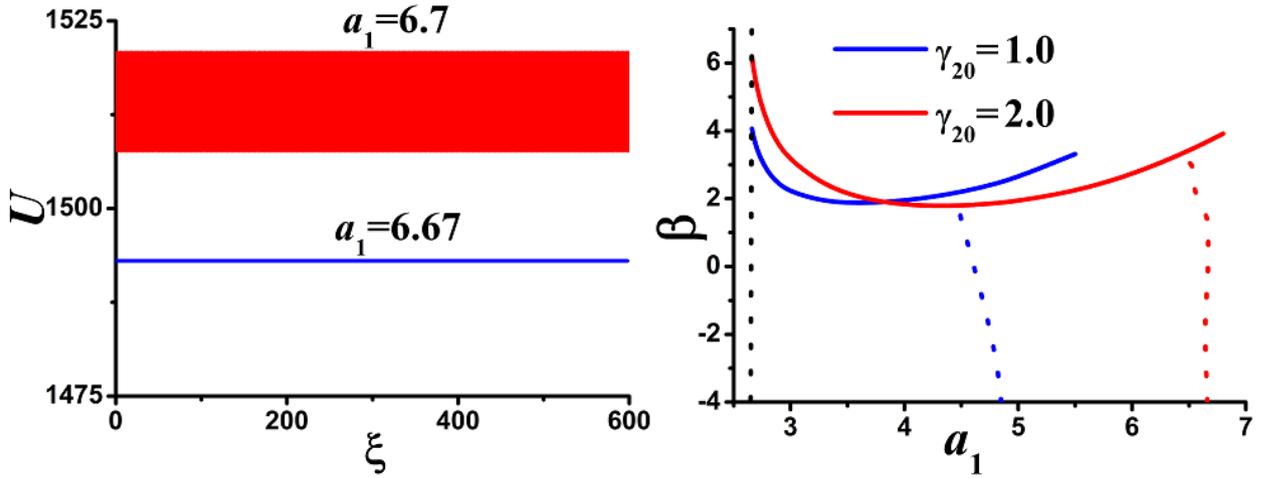

**Fig. 4.** Left: Evolution of total soliton power $U$ for the stable soliton at $a_1 = 6.67$, and for the unstable one at $a_1 = 6.7$. In both cases $W = 1.0$, $\gamma_{10} = 1.0$, $\gamma_{20} = 2.0$. The power of the unstable soliton oscillates in the course of the propagation, covering the red stripe. Right: The existence and stability domains for the solitons driven by the gain at the FF component for $\gamma_{20} = 1.0$ (blue lines) and $\gamma_{20} = 2.0$ (red lines). Solitons exist to the right of the vertical black dotted line and below the solid lines. Stability domains are located to the left of the dashed lines of the same color. In all cases $W = 1.0$, $\gamma_{10} = 1.0$. All quantities are plotted in dimensionless units.

## 4. Fundamental solitons supported by the localized gain applied at the second harmonic

Next, we aim to identify the soliton existence and stability domains for the dissipative system with the HS acting in the SH component. First, we looked for the radially symmetric solitons localized at the maximum of the gain profile. To address the generic case, we set $W = 1.0$, $\gamma_{10} = 1.0$, $\gamma_{20} = 1.0$. Dissipative-soliton solutions were found in a wide range of parameters $\beta$ and $a_2$. For $\beta \geq -3$, the solitons exist at $a_2 > a_{20} \approx 2.011$, where $a_{20}$ corresponds to the linear gain-guided mode in the SH equation, with $U_1 \to 0$ but $U_2 \neq 0$ (see left panels in Figs. 5 and 6). The finite value of $U_2$ is necessary for the compensation of the FF loss by the parametric gain generated by the $\chi^{(2)}$ term in the first equation of Eqs. (4). In this case only one radially symmetric dissipative solution exists, and dependence $U(a_2)$ is single-valued. For $-4.25 \leq \beta < -3$ several radially symmetric solutions coexist, see Fig. 7, and function $U(a_2)$ becomes multi-valued. We show only two stable or partially stable branches, while there is a totally unstable intermediate branch connecting upper and lower branches.

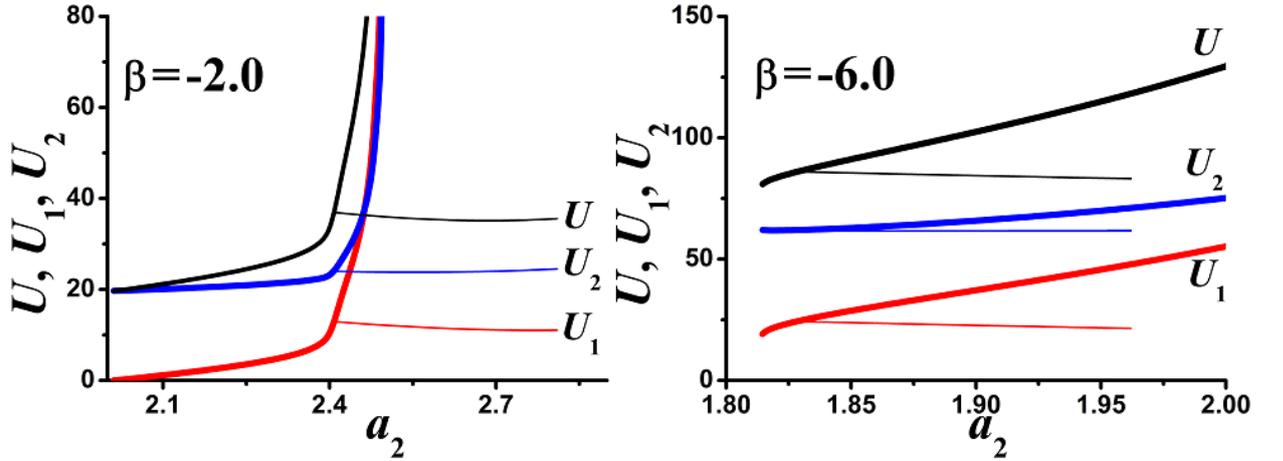

**Fig. 5**. The powers of the soliton components and the total power vs. the gain strength at $\beta = -2.0$ and $\beta = -6.0$ (the left and right panels, respectively). Thick and thin lines represent the radially symmetric and asymmetric (shifted) dissipative solitons, respectively. Symmetric solutions are stable before the appearance of the asymmetric solutions. Presented asymmetric solutions are fully stable. In all cases $W = 1.0$, $\gamma_{10} = \gamma_{20} = 1.0$. All quantities are plotted in dimensionless units.

For the lower branch, starting at $a_2 = a_{20}$, the range of the gain coefficient $a_2$ providing its existence becomes narrower with the decrease of $\beta$ and disappears at $\beta < -4.25$, see the left panels in Fig. 6. For $\beta < -4.25$ dependence $U(a_2)$ is again a single-valued function, and the solitons exist if the gain strength exceeds a critical value that depends on mismatch $\beta$. At the threshold points the tangential line to the $U(a_2)$ curve becomes vertical, while powers $U$, $U_1$, $U_2$ remain finite and nonvanishing, see the right panel in Fig. 5. For this upper branch the threshold value of the gain strength decreases with the decrease of $\beta$, but the minimum soliton power increases at the same time, see the left panel in Fig. 6. Thus, the upper branch exists for large absolute values of $\beta < 0$; as $\beta$ increases, a second lower branch appears (see Fig. 8), and then these branches merge. In all these cases, the solutions exist below some maximum value of the gain strength, which depends on the mismatch. If the gain is fixed, then the upper branch exists as a single one for $a_2 < 2.011$, several solutions may appear if $2.011 \leq a_2 \leq 2.15$ and two branches merge for $a_2 > 2.15$ (see the right panels in Fig. 6 and the left panel in Fig. 7).

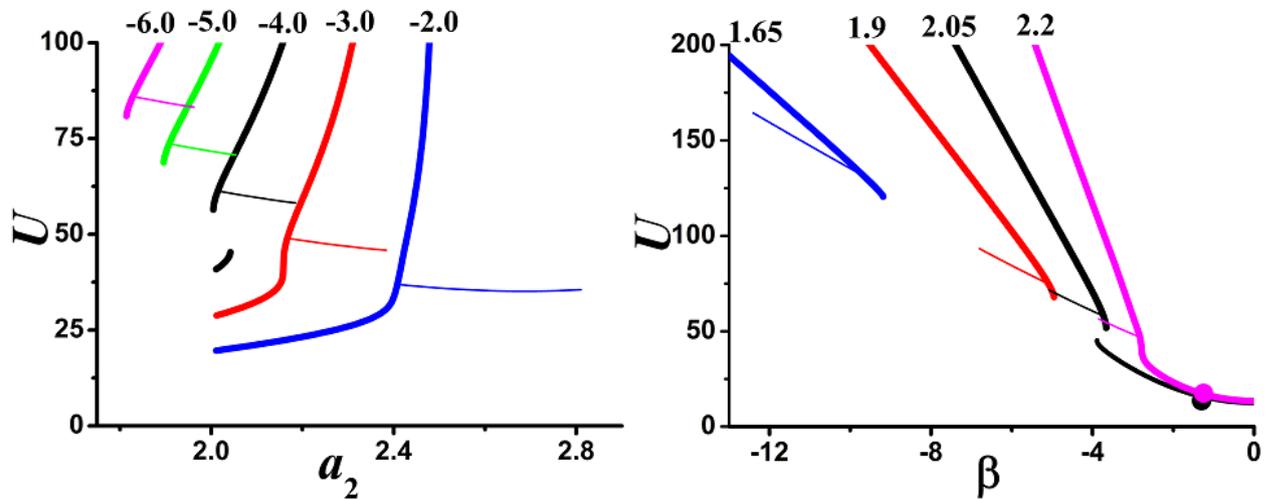

**Fig. 6.** The total power of the SH-driven solitons vs. the gain strength for different fixed values of the mismatch (the left panel), and the mismatch for different fixed values of the gain strength (the right panel). Thick and thin lines correspond, respectively, to the radially symmetric and asymmetric (shifted) solutions. Solid dots in the right panel denote boundaries of the stability domain for the lower branches of the same color. In both panels in the multi-valued case the lower branch is fully stable, while the upper one is stable up to the appearance of the asymmetric solutions. Asymmetric solutions presented here are fully stable. In all cases $W = 1.0$, $\gamma_{10} = \gamma_{20} = 1.0$. All quantities are plotted in dimensionless units.

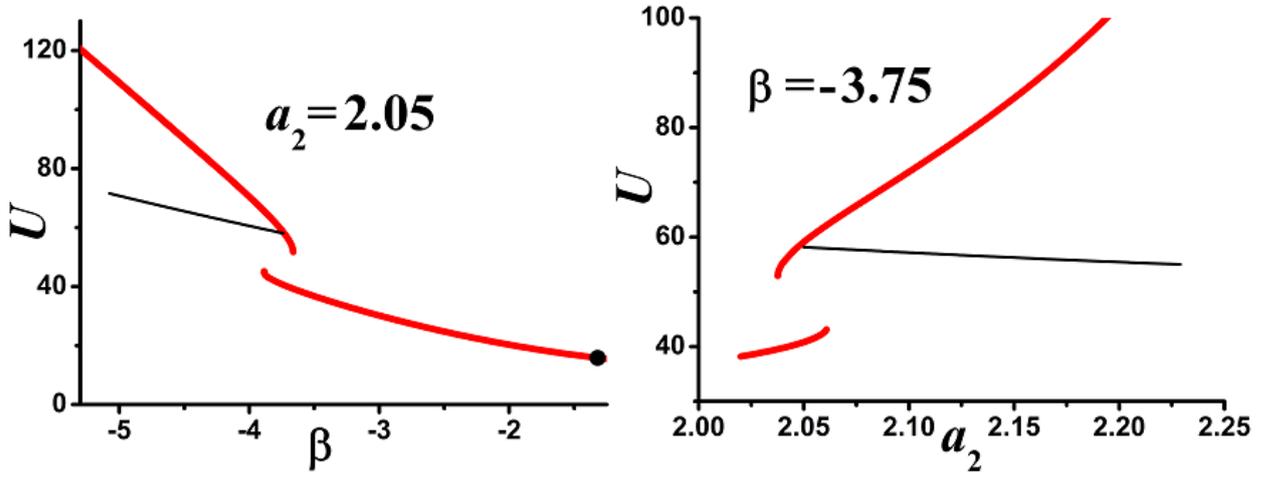

**Fig. 7.** The total power of the SH-driven dissipative solitons vs. the mismatch at $a_2 = 2.05$ (the left panel) and the SH gain strength at $\beta = -3.75$ (the right panel). Thick red and thin black lines correspond, respectively, to the radially symmetric and asymmetric (shifted) solutions. The black dot in the left panel denotes the boundary of the stability domain for the lower branch. In the right panel the lower branch is fully stable, while the upper branch is stable up to the appearance of the asymmetric solutions. Asymmetric solutions presented here are stable, in both panels.. In all cases, $\gamma_{10} = \gamma_{20} = 1.0$, $W = 1.0$. All quantities are plotted in dimensionless units.

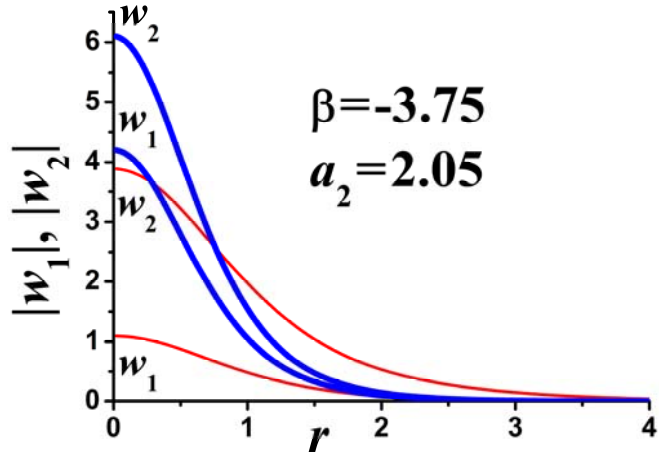

**Fig. 8.** Profiles of the soliton components from the upper branches (thick blue lines) and lower ones (thin red lines) at $\beta = -3.75$, $a_2 = 2.05$ $W = 1.0$, $\gamma_{10} = \gamma_{20} = 1.0$. All quantities are plotted in dimensionless units.

The stability of the radially symmetric solutions was studied by means of direct simulations of the perturbed evolution in the framework of Eq. (4). It was found that stable solutions exist solely for $\beta < 0$. The threshold value of $\beta$ slightly increases with the growth of the gain strength, see the solid dots in Fig. 6.

Next, at a fixed mismatch the radially symmetric solutions become unstable above a certain critical value of the gain strength. At this point, a stable stationary *asymmetric solution* appears, which is shifted off the central point, see right panels in Fig. 9. The shift increases with the growth of the gain strength. The range of the gain values maintaining stable asymmetric solutions widens with the decrease of the absolute values of negative $\beta$ (see thin lines in the left panel in Fig. 6). At large values of $|\beta|$ almost all symmetric solutions are unstable.

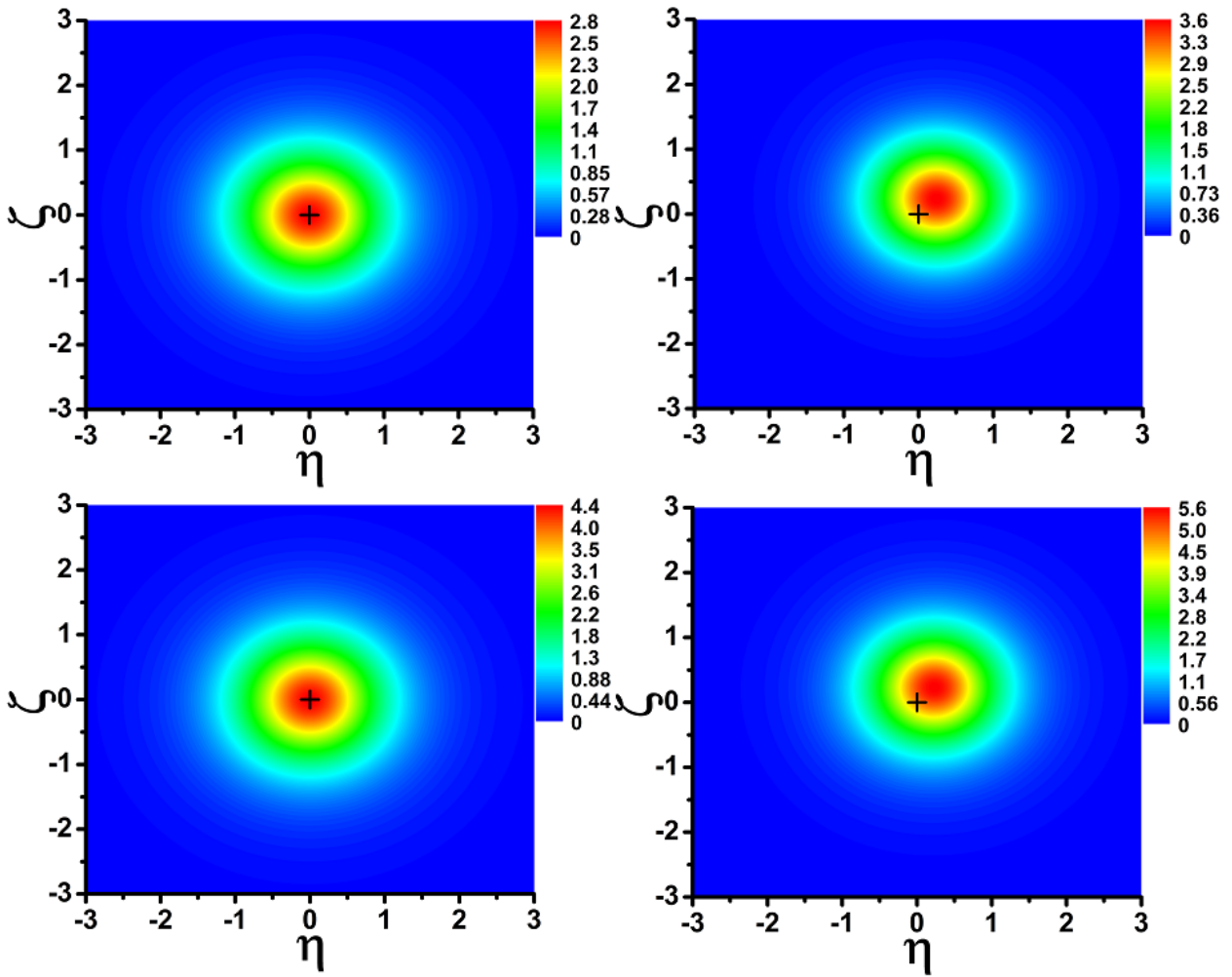

**Fig. 9.** The distribution of the absolute values of the FF and SH fields (the upper and bottom panels, respectively) in the SH-driven dissipative solitons at $\beta = -2.8$ and $\beta = -3.8$ (the symmetric and asymmetric ones, which are displayed, respectively, in the left and right panels) for $W = 1.0$, $\gamma_{20} = 1.0$, $\gamma_{10} = 1.0$, $a_2 = 2.2$. The black cross denotes the location of the HS's center. All quantities are plotted in dimensionless units.

If the gain is fixed, the asymmetric (shifted) solutions appear below a certain critical value of the mismatch. The mismatch range supporting the stable asymmetric solutions widens with the decrease of the gain strength (see the right panel in Fig. 6), while the size of the off-center shift becomes smaller (see Fig. 10): $\max(\Delta r / W) \approx 0.25$ for $a_1 = 1.65$, and $\max(\Delta r / W) \approx 0.48$ for $a_1 = 2.61$.

Note that, in the bistability domain, one may observe either two stable radially symmetric solutions. or a stable radially symmetric one belonging to the lower branch, coexisting with a stable asymmetric soliton, see Figs. 7 and 8. In this case, the type of generated soliton depends on parameters of the input beams. Namely, the generation of solitons from the upper branch or asymmetric solitons requires more powerful input Gaussian beams than the generation of solitons from the lower branch. The generation of shifted solitons is more effective if input beams are also shifted from the center of the gain region.

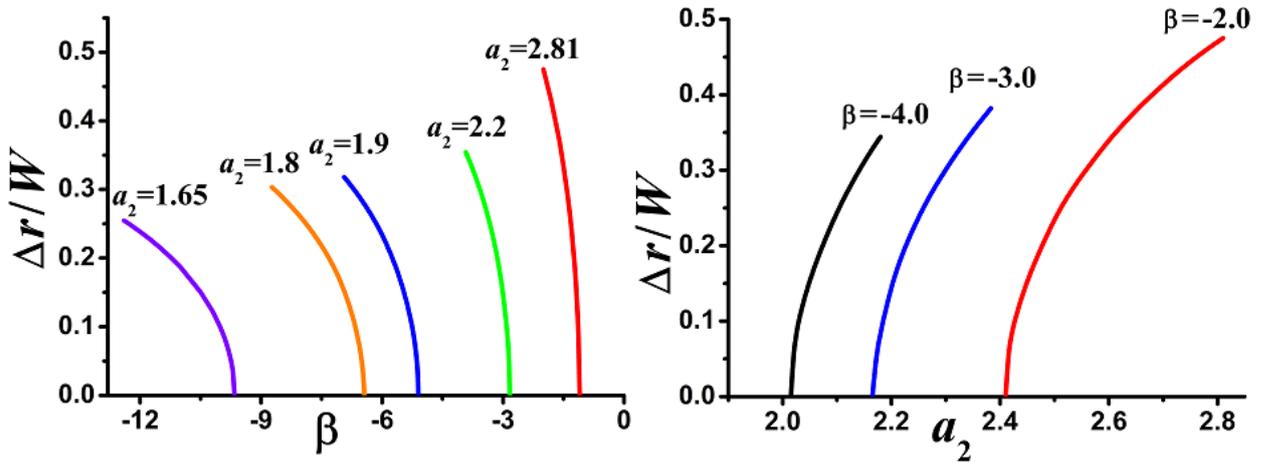

**Fig. 10.** The normalized shift of the stable asymmetric soliton vs. the mismatch for different values of the gain strength (the left panel), and vs. the gain strength for different mismatch values (right panel). In all cases $W = 1.0$, $\gamma_{10} = \gamma_{20} = 1.0$. All quantities are plotted in dimensionless units.

The results for the existence and stability domains of the fundamental solitons driven by the gain in the SH component are summarized in Fig. 11. The lower branch exists between solid blue lines. The upper branch exists between the red lines. The merged branch exists above the thick blue line and below the thin red one. The stability domain exists to the left of the green line. The symmetry breaking occurs above the solid black line, and the asymmetric solutions, shifted from the HS's center, are stable below the dotted black line.

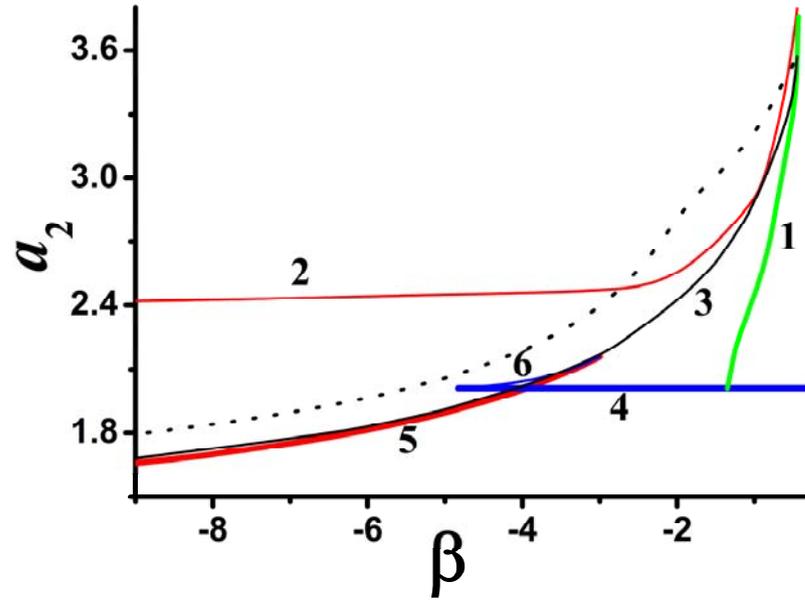

**Fig. 11.** The existence and stability domains for the solitons supported by the gain acting onto to the SH component for $W = 1.0$, $\gamma_{10} = 1.0$, $\gamma_{20} = 1.0$. The first symmetric branch (the lower ones, in terms of Fig. 7) exists between solid blue lines (marked "**4**" and "**6**"). The second symmetric branch (the upper one, in Fig. 7) exists between the red lines (marked "**2**" and "**5**"). The merged symmetric branch exists above the thick blue line (marked "**4**") and below the thin red one (marked "**2**"). The stability domain of symmetric solitons is one to the left of the green line (marked "**1**"). The off-center shifted asymmetric modes appear above the solid black line (marked "**3**"), and they are stable below the dotted black one. All quantities are plotted in dimensionless units.

Note that the results obtained for 2D fundamental solitons are very similar to finding for their 1D counterparts reported in Ref. [68]. However, vortex solitons considered in the next section demonstrate novel features that do not have any similarities in the 1D setting.

## 5. Vortex solitons supported by the ring-shaped gain profile applied at the FF component

A generic example of the ring-shaped gain profile driving the FF harmonic can be taken as per Eq. (2) with $W = R / 3 = 1.75$ at $\gamma_{10} = \gamma_{20} = 1.0$. The ring may be considered as a quasi-1D structure under the condition of $W \ll R$, which does not really hold in the present case, hence it does not amount to an effective 1D configuration.

Vortex-soliton solutions were searched as

$$q_1 = w_1(r)\exp\left(im\varphi + ib\xi\right), \ q_2 = w_2(r)\exp\left(2im\varphi + i(2b+\beta)\xi\right), \tag{10}$$

where $m$ is the topological charge. Note that the total angular momentum (8) of stationary states (10) is $M = mU$. Stationary solutions in the form of Eq. (10) were found by means of the relaxation technique, and existence domains for the radially-symmetric vortex solitons with different topological charges have been thus identified. Thus, it has been found that, for each value of integer winding number $m$ from 0 to 5, the solutions exist provided that, as above, the gain strength exceeds a respective critical value, $a_{10}(m)$. Also similar to the case of the fundamental mode (with $m = 0$), this value corresponds to the gain-guided linear mode ($U \to 0$ at $a_1 \to a_{10}$), and it practically does not depend on $m$, see the left panel in Fig. 12, being $a_{10} \approx 1.33$ for the present values of the parameters.

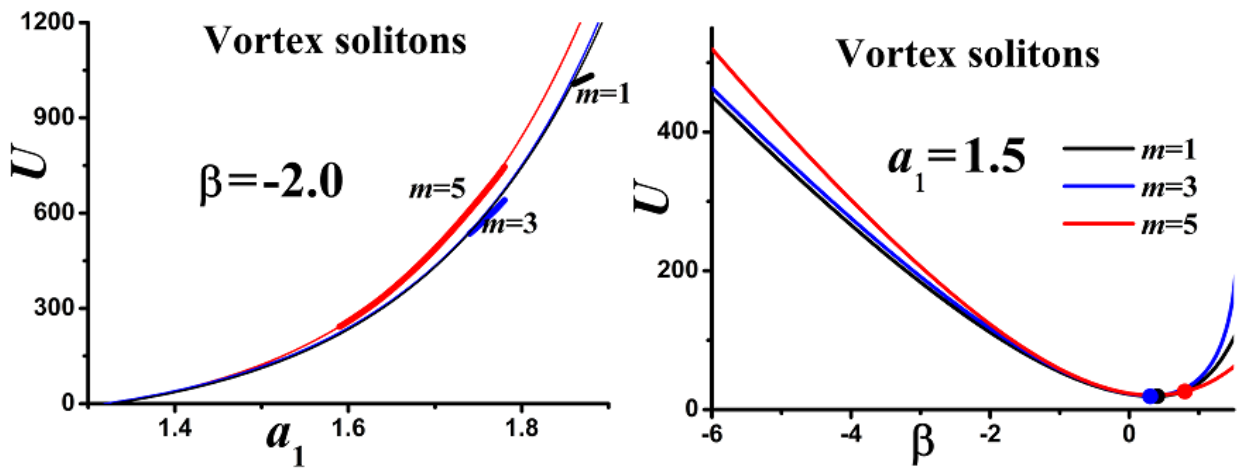

**Fig. 12.** The total vortex-soliton power vs. the gain strength at $\beta = -2.0$ (the left panel) and the mismatch at $a_1 = 1.5$ (the right panel) for different values of the winding number (topological charge). Thick lines in the left panel correspond to stable azimuthons replacing unstable axisymmetric vortex solitons. Solid dots in the right panel denote boundaries of the stability domain, the vortex solitons being stable to the left of these points. In all cases $W = 1.0$, $\gamma_{10} = \gamma_{20} = 1.0$. All quantities are plotted in dimensionless units.

Further, the existence and stability of such solutions was also examined running direct simulations Eq. (1). First, we obtained 2D vortex soliton profiles from inputs in the form of $q_1(\xi = 0) = A_1 r^m \exp\left(-(r/w_1)^2 + im\varphi\right)$, $q_2(\xi = 0) = 0$ ($w_1$ and $A_1$ are the width and amplitude of the FF component of the input beam). The stability of the solitons was then tested by adding perturbations to the generated profiles and simulating their subsequent evolution up to large distances. Stable solitons were found to keep their shape in the course of the propagation. The stability analysis has revealed that vortex solitons are stable if, chiefly, mismatch $\beta$ is negative; however, there is a narrow stability range for the vortices at $\beta > 0$, see the right panel in Fig. 12. At still larger values of $\beta$ the vortex-shaped input develops an azimuthal instability and spontaneously breaks into fragments (not shown here in detail). Characteristic shapes of the absolute value and phase of the fields for stable vortex solitons with $m = 1$ and $m = 5$ are displayed in Figs. 13 and 14. For considered parameters stable vortex solitons with $m > 5$ have not been found.

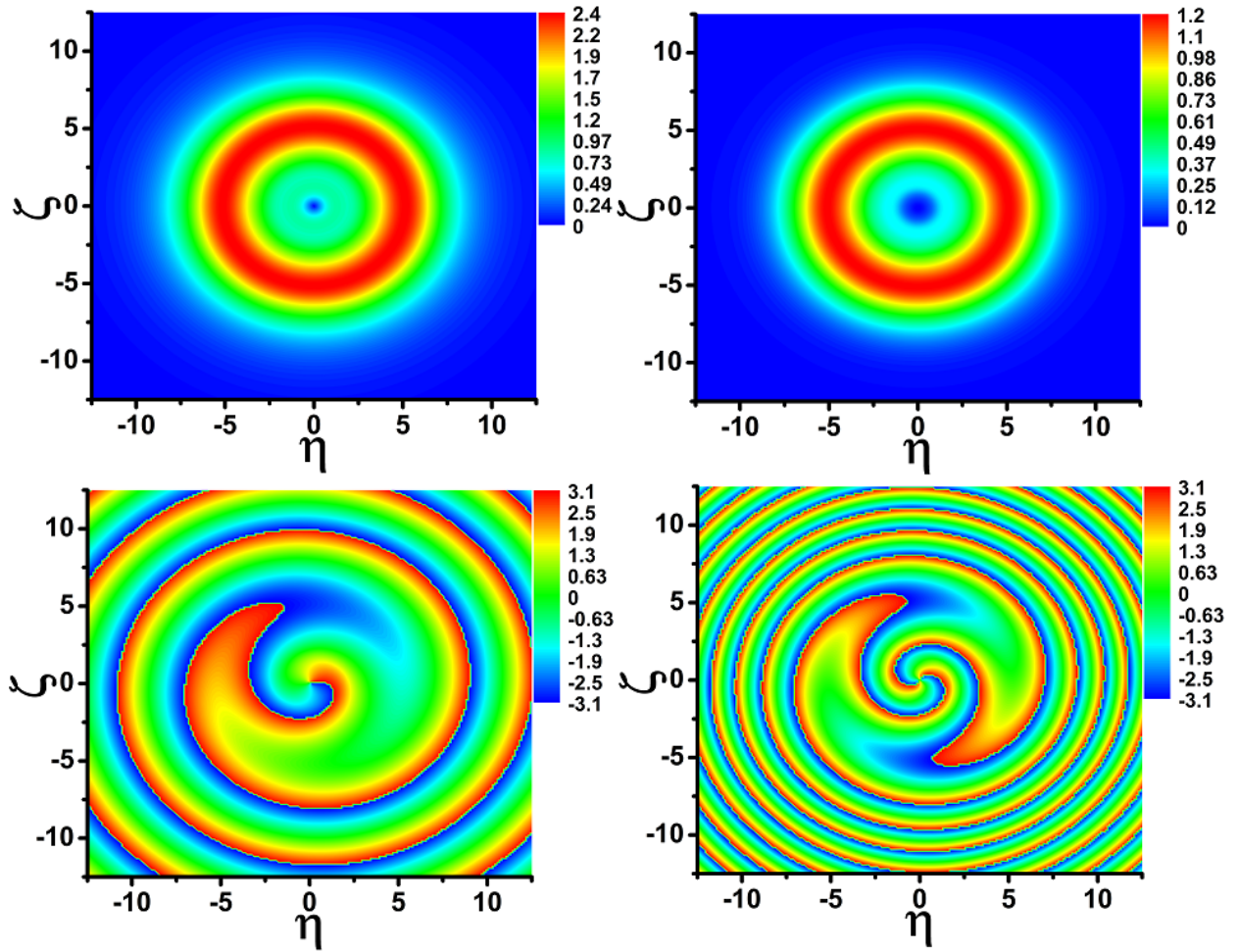

**Fig. 13.** Absolute values and phases (the upper and bottom panels, respectively) of the FF and SH components (the left and right panels, respectively) of the stable vortex soliton with $m = 1$ at $a_1 = 1.8$, $\beta = -2.0$, $R = 5.25$, $w = 1.75$, $\gamma_{20} = 1.0$, $\gamma_{10} = 1.0$. All quantities are plotted in dimensionless units.

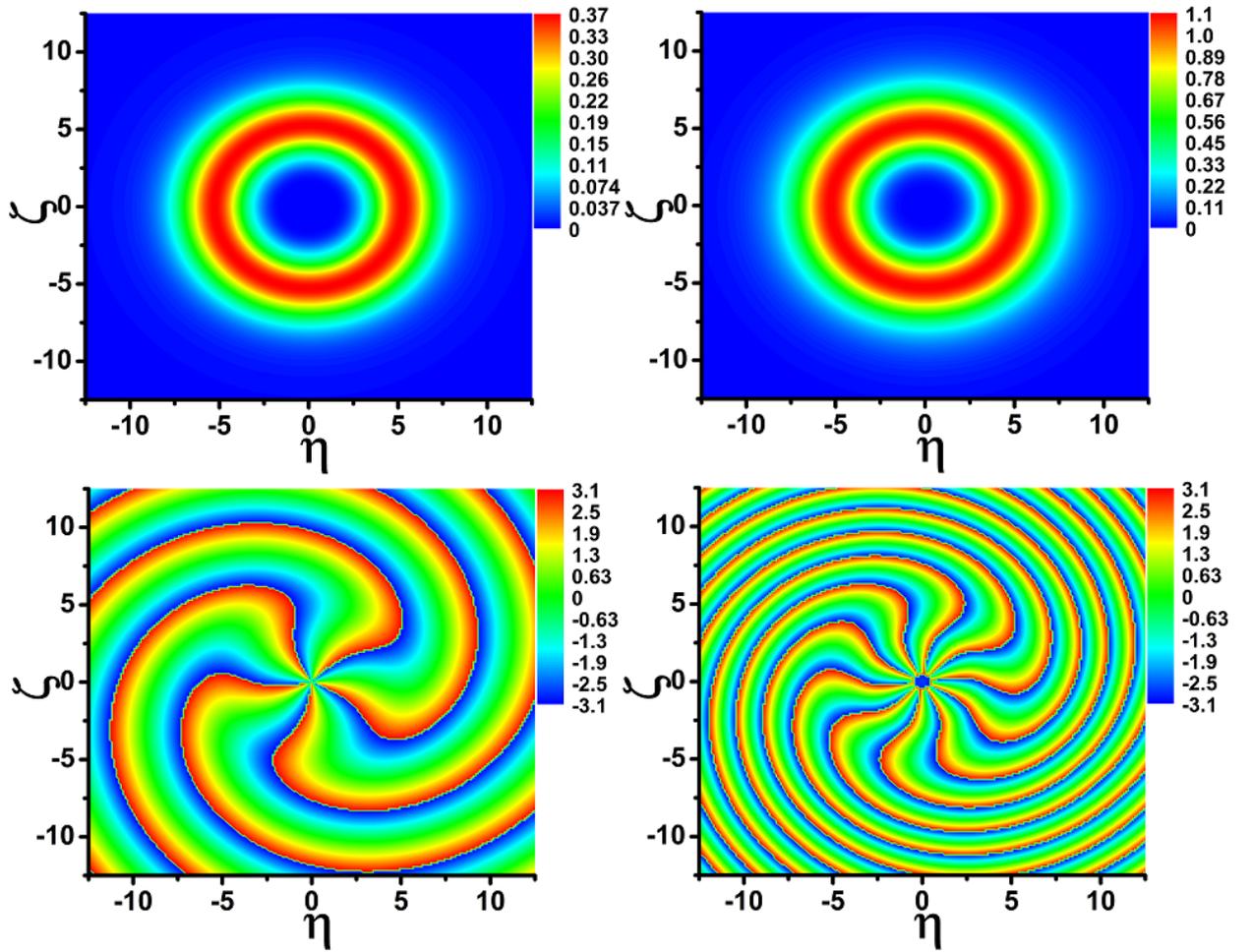

**Fig. 14.** Absolute values and phases (the top and bottom panels, respectively) of the FF and SH components (the left and right panels, respectively) for the stable vortex soliton with $m = 5$ at $a_1 = 1.5$, $\beta = -2.0$, $R = 5.25$, $W = 1.75$, $\gamma_{10} = \gamma_{20} = 1.0$. All quantities are plotted in dimensionless units.

We have also found that, at a fixed value of the mismatch, when the gain strength exceeds a critical level, the radially symmetric vortex soliton becomes unstable, but there appear stable azimuthally modulated solitons (*azimuthons* [72], represented by segments of the thick lines in the left panel in Fig. 12). The respective threshold values of the gain deceases with the growth of the winding number: at $\beta = -2.0$, $a_{1\text{th}} = 1.86$ for $m = 0$, $a_{1\text{th}} = 1.84$ for $m = 1$, $a_{1\text{th}} = 1.8$ for $m = 2$, $a_{1\text{th}} = 1.74$ for $m = 3$, $a_{1\text{th}} = 1.65$ for $m = 4$, and $a_{1\text{th}} = 1.58$ for $m = 5$. The azimuthal-modulation frequency depends on the winding number: we have observed 14 peaks in the modulated profile at $m = 0,1$, 13 peaks at $m = 2$, 12 peaks at $m = 3$, and 11 peaks at $m = 4$ and 5. The azimuthal modulation gets more pronounced upon further increase of the gain strength (see Fig. 15).

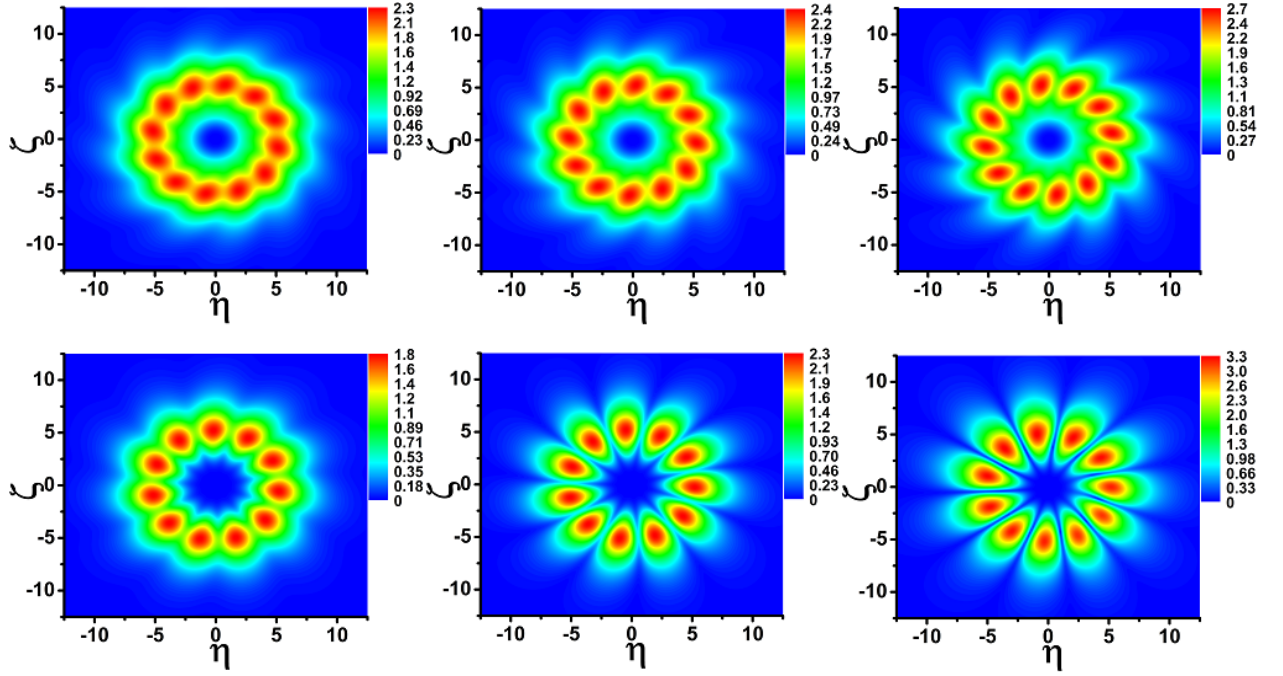

**Fig. 15.** Absolute values of the FF component of the stable vortex azimuthons with $m = 3$ and $5$ (the upper and bottom panels, respectively). The top row: $a_1 = 1.74$ (left), $a_1 = 1.75$ (middle), $a_1 = 1.78$ (right). The bottom row: $a_1 = 1.59$ (left), $a_1 = 1.65$ (middle), $a_1 = 1.78$ (right). In all cases $\beta = -2.0$, $R = 5.25$, $W = 1.75$, $\gamma_{10} = \gamma_{20} = 1.0$. All quantities are plotted in dimensionless units.

Azimuthons exhibit rotation in the course of the propagation. In particular, one can notice counterclockwise rotation of the azimuthon with $m = 5$ in Fig. 16.

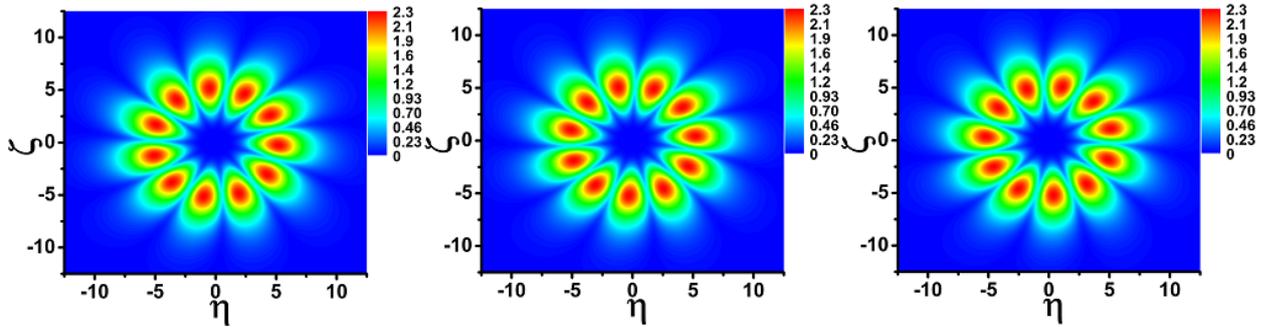

**Fig. 16.** The absolute value of the FF component of the azimuthon with $m = 5$ for $a_1 = 1.5$, $\beta = -2.0$, $R = 5.25$, $W = 1.75$, $\gamma_{10} = \gamma_{20} = 1.0$ at $\xi = 0$ (left panel), $\xi = 4$ (middle panel), $\xi = 8$ (right panel). All quantities are plotted in dimensionless units.

Stability of the azimuthons was checked by propagating them with added noise, in the framework of Eq. (1). Eventually, the vortex azimuthons become unstable above certain critical values of the gain strength.

Interestingly, at $a_1 > 1.78$ the unstable vortex azimuthons with $m = 3, 4, 5$ may spontaneously transform into a stable multipole (12-pole) soliton, see Fig. 17. For $\beta = -2.0$ stable 12-pole solitons were found in the interval of the gain strength

$$1.5 \le a_1 \le 2.025. \tag{11}$$

Their stability was also tested by simulating their propagation, with added noise, in the framework of Eq. (1). As the gain approaches the upper boundary of the stability range, the periodic structure of the multipole starts to distort, see the right column in Fig. 17. On the other hand, the 12-pole solitons themselves become unstable and spontaneously transform into axisymmetric vortex solitons (10) with $m = 5$ at $a_1 \le 1.5$, cf. Eq. (11). Interestingly, in contrast to the vortex azimuthons, the multipoles do not rotate.

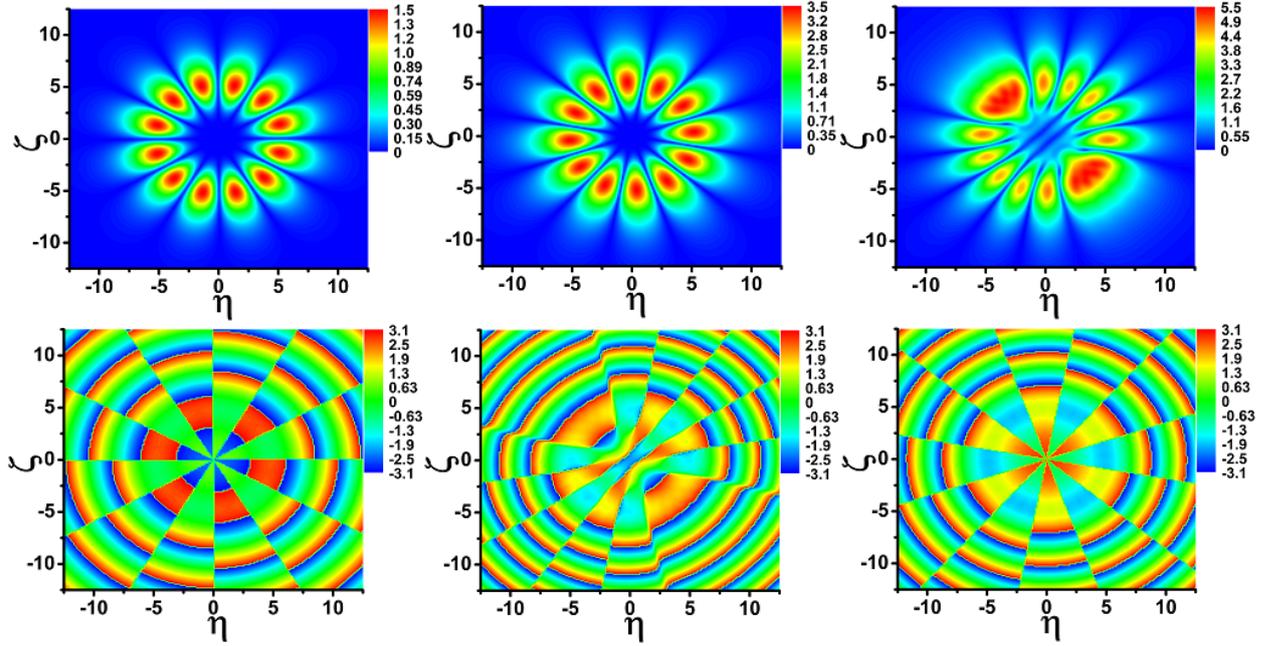

**Fig. 17.** The absolute value and phase (the top and bottom panels, respectively) of the FF component of the stable 12-pole soliton at $a_1 = 1.5$ (left), $a_1 = 1.8$ (middle), $a_1 = 2.0$ (right). In all cases $\beta = -2.0$, $R = 5.25$, $W = 1.75$, $\gamma_{10} = \gamma_{20} = 1.0$. All quantities are plotted in dimensionless units.

It is relevant to stress the difference of the two-component (FF-SH) vortex solitons from their single-component counterparts produced by the single CGL equation with the cubic nonlinearity, that were reported in Refs. [40] and [41]. As shown above, the FF-SH system of equations (1) reduces to the single cubic (3) in the cascading limit, corresponding to the large mismatch $|\beta|$, hence in this limit the two-component vortices are indeed similar to their single-component CGL counterparts. On the other hand, at moderate values of the mismatch the vortices form a broad family, intrinsically parameterized by coefficient β, as shown in the right panel of Fig. 12. The family includes, in particular, both stable and unstable vortices, and admits the variation of the total power in broad limits. Furthermore, unlike the case of the single GGL equation, unstable FF-SH vortices may spontaneously transform into the rotating azimuthons, as shown above in Figs. 15 and 16.

Further, we consider an effect of elliptical deformation of the gain profile on the vortex solitons. To this end, the axisymmetric profile $\gamma_1(r)$, taken as per Eq. (2), was replaced by $\gamma_1\left(\sqrt{\varepsilon^2\eta^2 + \zeta^2}\right)$, where $\varepsilon$ determines the eccentricity of the deformed profile, $e = \sqrt{1 - \varepsilon^2}$. It has thus been found that, similar to Ref. [45], the single high-winding-number phase dislocation (the vortex' pivot) splits into a set of spatially separated unitary dislocations, see Fig. 18.

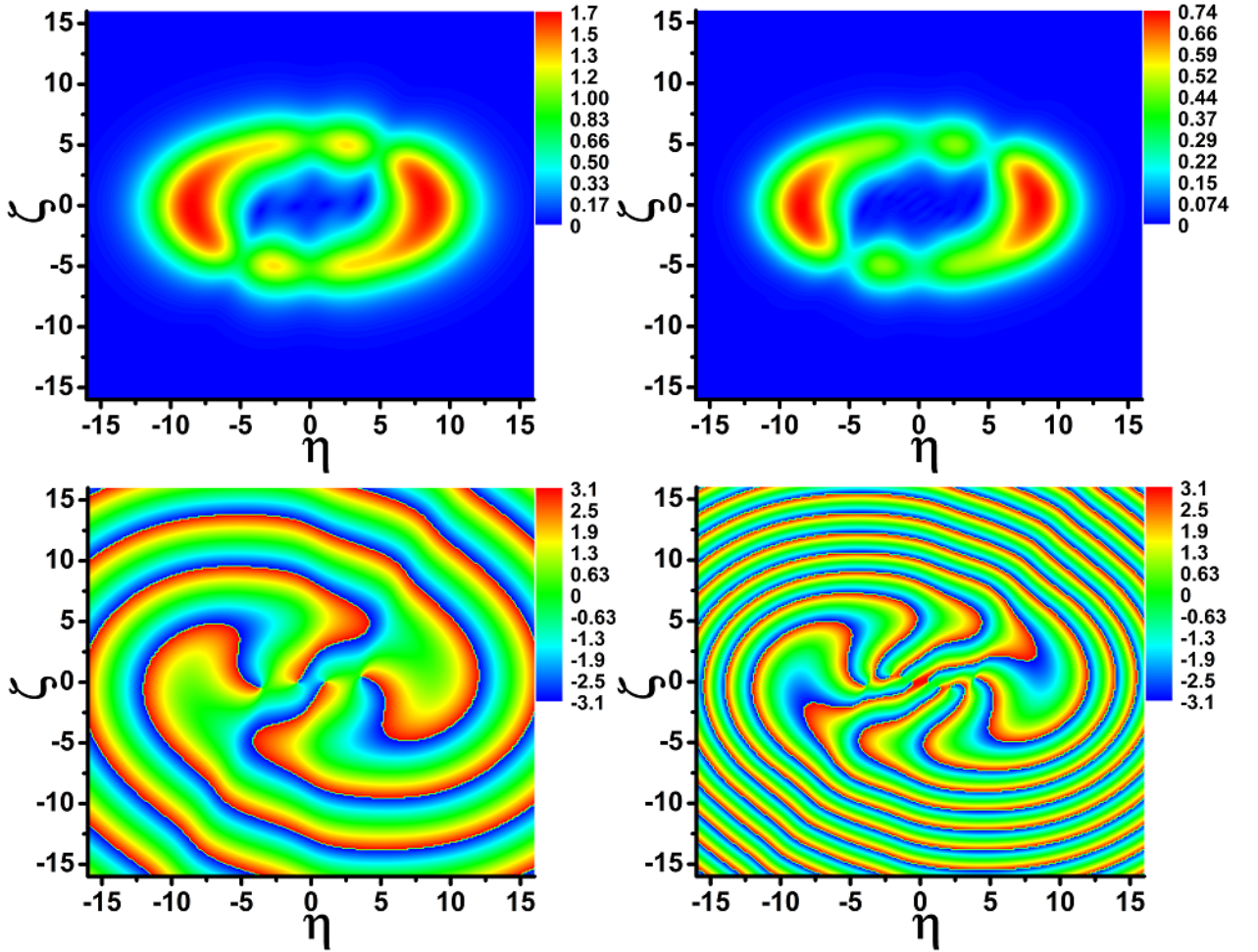

**Fig. 18.** Absolute values and phases (the top and bottom panels, respectively) of the FF and SH components (the left and right panels, respectively) of the stable vortex soliton with $m = 4$ at $a_1 = 1.5$, $\beta = -2.0$, $R = 5.25$, $W = 1.75$, $\gamma_{10} = \gamma_{20} = 1.0$, $\varepsilon = 0.6$. All quantities are plotted in dimensionless units.

Lastly, it is relevant to mention that the numerical analysis cannot produce stable vortex solitons supported by the ring-shaped gain profile acting onto the SH component. This fact may be explained by the propensity of the SH field to the parametric instability against small perturbations represented by the FF component [57-59].

## 6. Conclusion

In this work, we have demonstrated that the possibility to maintain stable 2D dissipative solitons by localized linear gain, applied to a lossy medium in a confined area, works well in the system with the $\chi^{(2)}$ (second-harmonic-generating) nonlinearity; previously, this possibility was addressed only in the 1D setting. The systematic numerical analysis has revealed the existence and stability regions for the zero-vorticity solitons supported by the Gaussian-shaped HS ("hot spot") acting in the FF (fundamental-frequency) or SH (second-harmonic) components. Along with the fundamental solitons, additional ones may be stable – in particular, asymmetric ones, which feature spontaneous off-center shift. The ring-shaped HS acting in the FF component supports stable vortex solitons, with winding numbers up to 5. Unstable vortex solitons transform into rotating azimuthons or ring-shaped multipoles.

As an extension of the work, it may be interesting to consider the non-degenerate system with two FF components.

## Acknowledgments

VEL and OVB acknowledge the personal support from the Foundation for the Advancement of Theoretical Physics and Mathematics "BASIS". The work of BAM was supported, in part, by the Israel Science Foundation through grant No. 1286/17.